\documentclass[11pt]{article}
\usepackage{cite}
\usepackage{wrapfig}
\usepackage{graphicx}
\usepackage{amssymb}
\usepackage{amsfonts}
\usepackage{amsmath}
\usepackage{longtable}
\usepackage{rotating}
\usepackage{lscape}
\usepackage{epsfig}
\usepackage{multirow}
\usepackage{color}
\usepackage{authblk}
\begin{document}
\title{\textbf{Integrable Cosmological Models with\\ an Arbitrary Number of Scalar Fields}}

\author[1]{V.R.~Ivanov\footnote{E-mail: vsvd.ivanov@gmail.com}}
\author[2]{S.Yu.~Vernov\footnote{E-mail: svernov@theory.sinp.msu.ru}}

\affil[1]{\,Physics Department, Lomonosov Moscow State University,
 Leninskie Gory~1, 119991, Moscow, Russia}

\affil[2]{\,Skobeltsyn Institute of Nuclear Physics, Lomonosov Moscow State University,
 Leninskie Gory~1, 119991, Moscow, Russia}

\date{ \ }

\maketitle
\begin{abstract}
We consider cosmological models with an arbitrary number of scalar fields nonminimally coupled to gravity and construct new integrable cosmological models. In the constructed models, the Ricci scalar is an integral of motion irrespectively of the type of metric. The general solutions of evolution equations in the spatially flat FLRW metric have been found for models with the quartic potentials.
\end{abstract}
\vspace*{6pt}

\noindent
PACS: 98.80.Cq; 04.50.Kd; 04.20.Jb


\label{sec:intro}
\section{Introduction}

Recent observations show with high accuracy that the Universe is isotropic and homogeneous at large scales.
For this reason, cosmological models with scalar fields play a central role in describing the global evolution
of the Universe. There are a lot of models with one or a few scalar fields that describe different epochs of the Universe evolution.

The Universe is spatially flat on large scales, so, the spatially flat Fried\-mann--Lema\^{i}tre--Robert\-son--Walker (FLRW) metric with
\begin{equation}
\label{metricFLRW}
ds^2 = { } - N^2(\tau)d\tau^2 + a^2(\tau)\left(dx^2 + dy^2 + dz^2\right),
\end{equation}
where $N(\tau)$ is the lapse function and $a(\tau)$ is the scale function of the parametric time $\tau$, can be used to describe  the global Universe evolution at the background level.

In FLRW metric, the Einstein's equations reduce to a system of ordinary differential equations, but this system is integrable only for a few specific forms of the potential.
The problem of constructing integrable cosmological models is actively investigated, but even taking into account single-field cosmological models, the number of integrable models is not large~\cite{Salopek:1990jq,Maciejewski:2008hj,Bars:2010zh,Fre:2013vza,Kamenshchik:2013dga,Boisseau:2015hqa,Kamenshchik:2015cla,Kamenshchik:2017ojc,Giacomini:2021xsx}. In the case of a few scalar fields, even construction of cosmological models with exact particular solutions is not a trivial task~\cite{Arefeva:2005mka,Vernov:2006dm,Andrianov:2007ua,Arefeva:2009tkq,Paliathanasis:2018vru,Chervon:2019nwq,Faraoni:2021opj,Fomin:2021snm,Mitsopoulos:2021qpu,Russo:2023nir}.

There are a few methods for integrating equations of evolution. The integrability of many cosmological models has been proven by solving evolution equations with suitable parametric time~\cite{Fre:2013vza}, but it is not clear how to find this parametric time. A method for constructing integrable models with non-minimally coupled scalar fields by using the relation between the Jordan and Einstein frames has been investigated for single-field models in Refs.~\cite{Kamenshchik:2013dga,Kamenshchik:2015cla,Kamenshchik:2016atu}. The integrability of cosmological models with non-minimal coupling can be more apparent than the integrability of the corresponding models in the Einstein frame~\cite{Kamenshchik:2015cla}.

The goal of this paper is to construct new integrable cosmological models with a few scalar fields that have the Ricci scalar as an integral of motion. One field cosmological model with such a property has been constructed in Ref.~\cite{Boisseau:2015hqa}. We construct models with an arbitrary number of non-minimally coupled scalar fields. For constructed models, we find the general solutions of evolution equations in the spatially flat FLRW metric~(\ref{metricFLRW}).

\section{Model with $N$ scalar fields and nonminimal coupling}

Let us consider the model, described by the following action
\begin{equation}
S = \int d^4 x \sqrt{-g} \left(U\left(\phi^A\right) R - \frac12 g^{\mu \nu} \delta_{AB}\partial_\mu \phi^A \partial_\nu \phi^B - V\left(\phi^A\right)\right),
\end{equation}
where capital Latin indices enumerate fields, $A=1,2,\dots N$, and are lowered and raised with the Kronecker symbol $\delta_{AB}$. The functions $U(\phi^A)$ and $V(\phi^A)$ are twice-differentiable functions, $R$ is the Ricci scalar.

The corresponding Einstein's equations are~\cite{Kamenshchik:2017ojc}
\begin{equation}
\label{Equmunu}
U\left(R_{\mu \nu} - \frac12 g_{\mu \nu} R\right) = \nabla_\mu \nabla_\nu \, U - g_{\mu \nu} \Box \, U + \frac12 T_{\mu \nu},
\end{equation}
where
\begin{equation}
T_{\mu \nu} = \delta_{AB}\partial_\mu \phi^A \partial_\nu \phi^B -  g_{\mu \nu} \left(\frac12 g^{\alpha \beta} \delta_{AB} \partial_\alpha \phi^A \partial_\beta \phi^B + V\right).
\end{equation}

The field equations have the following form
\begin{equation}
\label{equphi}
\Box \phi^A =  V_{,\phi^A} - R U_{,\phi^A},
\end{equation}
where $F_{,\phi^A}=\partial F/\partial \phi^A$ for any function $F$.

From Eq.~(\ref{Equmunu}), we get
\begin{equation}
\label{equR}
    3\Box U -UR=\frac12 g^{\mu\nu}T_{\mu\nu}={ } -\frac12 g^{\mu\nu}\delta_{AB}\partial_\mu\phi^A\partial_\nu\phi^B - 2V.
\end{equation}
Writing $\Box U$ explicitly,
\begin{equation}\label{BoxU}
    \Box{U}=U_{,\phi^A}\Box\phi^A+U_{,\phi^A\phi^B}g^{\alpha\beta}\partial_\alpha \phi^A \partial_\beta \phi^B,
\end{equation}
we obtain from Eq.~(\ref{equR})
\begin{equation}
\label{equRUf}
    \left(3 U_{,\phi^A}^2+U\right)R -\left[3 U_{,\phi^A\phi^B}+\frac{\delta_{AB}}{2}\right]g^{\alpha\beta}\partial_\alpha \phi^A \partial_\beta \phi^B=3U_{,\phi^A}V_{,\phi^A}+2V.
\end{equation}
It is easy to see that for the function
\begin{equation}
\label{U_choice}
U = U_0 - \frac{\delta_{AB}}{12}\left(\phi^A - \phi_0^A\right)\left(\phi^B - \phi_0^B\right),
\end{equation}
where $\phi_0^A$ is an arbitrary constant vector, Eq.~(\ref{equRUf}) takes the following form:
\begin{equation}
\label{diffequV}
    \frac{1}{2}\left(\phi^A - \phi_0^A\right)V_{,\phi^A}-2V+U_0R=0.
\end{equation}
If $R$ is a constant, $R=R_0$, then Eq.~(\ref{diffequV}) is a linear first-order partial differential equation with $V(\phi^A)$ as an unknown function and the components of $\phi^A$ as independent variables. Equation~(\ref{diffequV}) has the general solution
\begin{equation}
\label{V_choice}
V = \frac{R_0 U_0}{2} + \left(\phi^1 - \phi_0^1\right)^4 f\left(\frac{\phi^2 - \phi_0^2}{\phi^1 - \phi_0^1}, \frac{\phi^3 - \phi_0^3}{\phi^1 - \phi_0^1}, \dots, \frac{\phi^N - \phi_0^N}{\phi^1 - \phi_0^1}\right),
\end{equation}
where $f$ is an arbitrary differentiable function.

So, we find that the Ricci scalar $R$ is the integral of motion for models with $N$ scalar fields and functions $U$ and $V$ given by Eqs.~(\ref{U_choice}) and (\ref{V_choice}), respectively.
Note that we do not specify the form of the metric. The vector $\phi_0^A$ can be put to zero without loss of generality.

We have considered the case of $N$ scalar fields with standard kinetic terms. To generalize this result to the case with a phantom scalar field $\phi^A$, it is enough to assume that the scalar field $\phi^A$ has the standard kinetic part, but is a purely imaginary function. Note that the function $U$ remains real in this case, and the potential $V$ is real for any real-valued function~$f$.

\section{General solutions in the FLRW metric}
In the FLRW metric (\ref{metricFLRW}),
Eqs.~(\ref{Equmunu}) and (\ref{equphi}) take the following form:
\begin{equation}
\label{G00_FLRW}
6 U h^2{} +6 h \dot{U}  =\frac12 \delta_{AB} \dot{\phi}^A \dot{\phi}^B + N^2V,
\end{equation}
\begin{equation}
\label{G11_FLRW}
{}-U\left(2\dot{h} + 3h^2\right) -2Uh\frac{\dot{N}}{N}= \ddot{U} + 2h\dot{U}-\dot{U}\frac{\dot{N}}{N} + \frac12\left(\frac12\delta_{AB} \dot{\phi}^A \dot{\phi}^B - N^2V\right),
\end{equation}
\begin{equation}
\label{fields_FLRW}
\ddot{\phi}^A + \left(3 h-\frac{\dot{N}}{N}\right) \dot{\phi}^A +V_{, \phi^A}N^2 - 6\left(\dot{h} + 2 h^2-h\frac{\dot{N}}{N}\right) U_{,\phi^A} = 0,
\end{equation}
where  $h(\tau)\equiv\dot{a} / a$, dots mean derivatives with respect to the parametric time $\tau$, and we assume that all fields are homogeneous.

For the function $U$ given by~(\ref{U_choice}),  the field equations~(\ref{fields_FLRW}) take the form
\begin{equation}
\label{equphiFLRW}
\ddot{\phi}^A + \left(3 h-\frac{\dot{N}}{N}\right) \dot{\phi}^A +V_{,\, \phi^A}N^2 + \left(\dot{h} + 2 h^2-h\frac{\dot{N}}{N}\right)\phi^A = 0
\end{equation}

To solve Eq.~(\ref{equphiFLRW}), we use the conformal time $\eta$, defined as $dt=a(\eta)d\eta$, hence, $N(\eta)=a(\eta)$. In terms of functions $y^A(\eta) \equiv a(\eta)\ \phi^A(\eta)$, the field equations (\ref{equphiFLRW}) are
\begin{equation}
\frac{d^2 y^A}{d \eta^2} + a^3  V_{,\phi^A}\left(\frac{y^1}{a}, \dots, \frac{y^N}{a}\right) = 0.
\end{equation}

Moreover, if we take $V$ as in~(\ref{V_choice}), we get
\begin{equation}
\label{y_system}
\frac{d^2 y^A}{d\eta^2} + \frac{\partial}{\partial y^A}V\left(y^1, \dots, y^N\right) = 0,
\end{equation}
where
\begin{equation*}
V\left(y^1, \dots, y^N\right)= \frac12 R_0 U_0+\left(y^1\right)^4f\left(\frac{y^2}{y^1}, \frac{y^3}{y^1},\dots, \frac{y^N}{y^1}\right).
\end{equation*}

The system (\ref{y_system}) has the first integral
\begin{equation}
\frac12 \delta_{AB} \frac{d y^A}{d \eta} \frac{d y^B}{d \eta} + V\left(y^1, \dots, y^N\right) = \frac12 R_0 U_0+ E,
\end{equation}
where $E$ is an integration constant.

For the previously chosen functions $U$ and $V$, the constraint equation~(\ref{G00_FLRW}) in the conformal time takes the following form
\begin{equation}
\label{Equ00eta}
\left(\frac{h}{a}\right)^2 = \frac{R_0}{12} + \frac{E}{6U_0 a^4}.
\end{equation}

Let's consider a particular kind of the potential $V$:
\begin{equation}
\label{pot_sum}
V(\phi^1, \dots, \phi^N) = \frac{R_0 U_0}{2} + \sum_{A=1}^{N}c_A \left(\phi^A\right)^4,
\end{equation}
where $c_A$ are constants.

With such a potential, system~(\ref{y_system}) is completely separable, and we have $N$ independent integrals of motion:
\begin{equation}
\label{y_system_fin}
\frac12 \left(\frac{d y^A}{d \eta}\right)^2 + c_A \left(y^A\right)^4 = C^A,
\end{equation}
where $C^A$ are integration constants, and the Einstein summation convention is not applied to $A$. We get that $E = \sum\limits_{A=1}^{N}C^A$, and
\begin{equation}
H^2(\tau)=\left(\frac{h}{a}\right)^2 = \frac{R_0}{12} + \frac{1}{6 U_0 a^4}\sum_{n=1}^{N}C^A.
\end{equation}
For the lapse function $N=1$, the condition $R=R_0$ gives the first order differential equation in the Hubble parameter $H(t)$,
\begin{equation}
\dot{H} + 2 H^2 = \frac{R_0}{6},
\end{equation}
that has the following general solution at $R_0>0$:
\begin{equation}
H(t)=\frac{\sqrt{3R_0}\left(\mathrm{e}^{2\sqrt {3R_0}t/3}+B\right)}{6\left(\mathrm{e}^{2\,
\sqrt {3R_0}t/3}-B\right)}\,,
\end{equation}
where $B$ is a constant.

There are several different solutions for $H$, $a$ and $\eta$, depending on the sign of $E$ and on whether or not $R_0$ is equal to zero.
Since the fields $y_n$ are most simply written as functions of $\eta$, it is helpful to write out $H$ and $a$ as functions of $\eta$ too, as well as to explicitly write the domain of $\eta$.

For $R_0>0$, we obtain
\begin{enumerate}
\item if $E > 0$, then
\begin{equation}
\begin{split}
&H(t) = \sqrt{\frac{R_0}{12}} \coth \left(\sqrt{\frac{R_0}{3}}(t - t_0)\right),\\
&a(t) =  a_0 \sqrt{\sinh\left(\sqrt{\frac{R_0}{3}}(t - t_0)\right)},
\end{split}
\end{equation}
\begin{equation*}
\eta(t) = \eta_0 + \frac{1}{a_0}\sqrt{\frac{3}{R_0}}F\left(\arccos\left(\frac{a_0^2 - a^2}{a_0^2 + a^2}\right)\left|\frac12\right.\right);
\end{equation*}
\begin{equation}
H(\eta) = \sqrt{\frac{R_0}{3}}\frac{\mathrm{dn}\left(\left.a_0\sqrt{\frac{R_0}{3}}\left(\eta-\eta_0\right)\,\right|\frac12\right)}{1 - \mathrm{cn}\left(\left.a_0\sqrt{\frac{R_0}{3}}\left(\eta-\eta_0\right)\,\right|\frac12\right)},
\end{equation}
\begin{equation}
a(\eta) =  \frac{a_0\, \mathrm{sn}\left(\left.a_0\sqrt{\frac{R_0}{3}}\left(\eta-\eta_0\right)\,\right|\frac12\right)}{1+\mathrm{cn}\left(\left.a_0\sqrt{\frac{R_0}{3}}\left(\eta-\eta_0\right)\,\right|\frac12\right)},
\quad a_0 = \left(\frac{2E}{U_0 R_0}\right)^{\frac14},
\end{equation}
\begin{equation*}
a_0\sqrt{\frac{R_0}{12}}\left(\eta-\eta_0\right) \in \left(0,  K\left(\frac12\right)\right);
\end{equation*}

\item  if $E < 0$, then
\begin{equation}
\label{bounce}
\begin{split}
&H(t) = \sqrt{\frac{R_0}{12}} \tanh \left(\sqrt{\frac{R_0}{3}}(t - t_0)\right),\\
&a(t) =  a_0 \sqrt{\cosh\left(\sqrt{\frac{R_0}{3}}(t - t_0)\right)},
\end{split}
\end{equation}
\begin{equation*}
\eta(t) = \eta_0 + \frac{1}{a_0}\sqrt{\frac{6}{R_0}}F\left(\arcsin\left(\sqrt{\frac{a^2 - a_0^2}{a^2}}\right)\left|\frac12\right.\right);
\end{equation*}
\begin{equation}
a(\eta) =  \frac{a_0}{\mathrm{cn}\left(\left.a_0\sqrt{\frac{R_0}{6}}\left(\eta-\eta_0\right)\,\right|\frac12\right)},\quad a_0 = \left(\frac{-2E}{U_0 R_0}\right)^{\frac14},
\end{equation}
\begin{equation}
H(\eta) = \sqrt{\frac{R_0}{6}}\mathrm{sn}\left(\left.a_0\sqrt{\frac{R_0}{6}}\left(\eta-\eta_0\right)\,\right|\frac12\right)\mathrm{dn}\left(\left.a_0\sqrt{\frac{R_0}{6}}\left(\eta-\eta_0\right)\,\right|\frac12\right),
\end{equation}
\begin{equation*}
a_0\sqrt{\frac{R_0}{6}}\left(\eta-\eta_0\right) \in \left(-K\left(\frac12\right), K\left(\frac12\right)\right);
\end{equation*}

\item if $E = 0$, then
\begin{equation}
H(t) = \pm\sqrt{\frac{R_0}{12}},\qquad a(t) = a_0 e^{\pm\sqrt{R_0/12}(t-t_0)},
\end{equation}
\begin{equation*}
\eta(t) = \eta_0 \pm \frac{1}{a_0}\sqrt{\frac{12}{R_0}}\left(1 - e^{\mp \sqrt{R_0/12}(t-t_0)}\right) =\eta_0 \pm \frac{1}{a_0}\sqrt{\frac{12}{R_0}}\left(1 -\frac{a_0}{a}\right);
\end{equation*}
\begin{equation}
H(\eta) = \pm\sqrt{\frac{R_0}{12}}, \qquad a(\eta) = \frac{a_0}{1 \mp a_0 \sqrt{R_0/12}\left(\eta-\eta_0\right)},
\end{equation}
\begin{equation*}
\pm a_0\sqrt{\frac{R_0}{12}}\left(\eta-\eta_0\right) \in \left(-\infty, 1\right);
\end{equation*}

\item if $R_0 = 0$ and $E > 0$, then
\begin{equation}
H(t) = \frac{1}{2(t-t_0)},
\quad a(t) = a_0 \sqrt{t - t_0},
\end{equation}
\begin{equation*}
\eta(t) = \eta_0 + \frac{2}{a_0}\sqrt{t-t_0} = \eta_0 + \frac{2}{a_0^2}a.
\end{equation*}
\begin{equation}
H(\eta) = \frac{2}{a_0^2(\eta-\eta_0)^2},\quad a(\eta) = \frac12 a_0^2\left(\eta-\eta_0\right),\quad a_0 = \left(\frac{2 E}{3 U_0}\right)^{\frac14},
\end{equation}
\begin{equation*}
\eta - \eta_0 \in \left(0, +\infty\right).
\end{equation*}
\end{enumerate}
Here $F$ is the incomplete elliptic integral of the first kind, defined as
\begin{equation}
F\left(\left.\varphi\,\right|m\right) = \int\limits_0^\varphi \frac{d\theta}{\sqrt{1 - m \sin^2 \theta}},
\end{equation}
$K(m) \equiv F(\pi / 2 \left| m\right.)$ is the complete elliptic integral of the first kind, and also the Jacobi elliptic functions $\mathrm{sn}$, $\mathrm{cn}$ and $\mathrm{dn}$ were introduced. By definition, if $u = F\left(\left.\varphi\,\right|m\right)$, then
\begin{equation}
\begin{array}{ll}
\mathrm{am} \left(\left.u\,\right|m\right) = \varphi, &\qquad \mathrm{dn}\left(\left.u\,\right|m\right) = \frac{d}{du} \mathrm{am} \left(\left.u\,\right|m\right),\\
\mathrm{sn}\left(\left.u\,\right|m\right) = \sin \left(\mathrm{am} \left(\left.u\,\right|m\right)\right),& \qquad
\mathrm{cn}\left(\left.u\,\right|m\right) = \cos \left(\mathrm{am} \left(\left.u\,\right|m\right)\right).\\
\end{array}
\end{equation}

The form of solutions of~(\ref{y_system_fin}) depends on the signs of $c_A$ and $C^A$:
\begin{enumerate}
\item $C^A > 0, \quad c_A > 0$:
\begin{equation*}
y^A(\eta) = \left(\frac{C^A}{c_A}\right)^\frac14 \mathrm{cn}\left.\left(\pm 2\left(C^A c_A\right)^\frac14 (\eta - \eta_i) + \mathrm{cn}^{-1}\left(u_A\left|\frac12\right.\right)\right| \frac12\right);
\end{equation*}
\item $C^A < 0, \quad c_A < 0$:
\begin{equation*}
y^A(\eta) = \frac{\left(\frac{C^A}{c_A}\right)^\frac14}{\mathrm{cn}\left.\left(\pm 2\left(C^A c_A\right)^\frac14 (\eta - \eta_i) + \mathrm{cn}^{-1}\left.\left(\frac{1}{u_A}\right|\frac12\right)\right| \frac12\right)};
\end{equation*}
\item $C^A > 0, \quad c_A < 0$:
\begin{equation*}
y^A(\eta) = \left|\frac{C^A}{c_A}\right|^\frac14\frac{\mathrm{sn}\left(\pm\left.2\sqrt{2}\left|C^A c_A\right|^\frac14 (\eta - \eta_i) + \mathrm{sn}^{-1}\left(\left.\frac{2u_A}{1+u_A^2}\right|\frac12\right)\right| \frac12\right)}{1+\mathrm{cn}\left(\pm\left.2\sqrt{2}\left|C^A c_A\right|^\frac14 (\eta - \eta_i) + \mathrm{cn}^{-1}\left(\left.\frac{1-u_A^2}{1+u_A^2}\right|\frac12\right)\right| \frac12\right)}.
\end{equation*}

Constants $u_A = |c_A/C_A|^{1/4} y^A(\eta_i)$ and the choice of sign in ``$\pm$'' are determined by initial conditions set at $\eta=\eta_i$. Note that no summation over $A$ is assumed.

\item For $c_A = 0$, $C^A$ has to be non-negative, and we have
\begin{equation*}
y^A(\eta) = y^A(\eta_i) + \frac{d y^A}{d \eta}(\eta_i) (\eta - \eta_i).
\end{equation*}

\item There are no solutions when $C^A < 0$ and $c_A \geqslant 0$.
\end{enumerate}

Note that the bounce solution (\ref{bounce}) exists only if at least one of $c_A<0$ and, therefore, the potential $V$ is not bounded from below.

\section{Conclusions}

In this paper, new integrable models with $N$ scalar fields non-minimally coupled to gravity have been constructed. We have shown that the Ricci scalar is an integral of motion if the functions $U$ and $V$ are given by Eqs.~(\ref{U_choice}) and (\ref{V_choice}). In the spatially flat FLRW metric, the Hubble parameter can be presented in terms of hyperbolic functions of the cosmic time. The behaviour of the scalar fields have been described in terms of the elliptic functions of the conformal time.

The corresponding one field model proposed in Ref.~\cite{Boisseau:2015hqa} can be transformed to the integrable model with a scalar field minimally coupled to gravity~\cite{Kamenshchik:2015cla}.
In general case, it is not possible to transform models with a few scalar fields non-minimally coupled to gravity into models with minimally coupled scalar fields with standard kinetic terms~\cite{Kaiser:2010ps}. After the metric transformation, one obtains models with a non-standard kinetic part, the so-called chiral cosmological models~\cite{Chervon:1995jx,Chervon:2013btx,Chervon:2019nwq}.

The simplest case of an integrable chiral cosmological model is the two-field model with a constant potential. The absence of the scalar field dependence of the potential allows one to obtain a description of the behaviour of the Hubble parameter in the form of  hyperbolic functions~\cite{Ivanov:2021ily}.

New integrable models with minimally coupled scalar fields can be obtained by conformal metric transformation of integrable models with non-minimally coupled scalar fields~\cite{Ivanov:2024nnd}. Another possible way to generalize the obtained results is to get the general solutions of integrable FLRW models in open and closed FLRW metrics~\cite{Kamenshchik:2015cla,Bars:2011mh} or in the Bianchi I metric~\cite{Kamenshchik:2017ojc,Kamenshchik:2016atu,Ivanov:2023uvh}.


\label{sec:funding}
\section*{Funding}
V.R.I. is supported by the ``BASIS'' Foundation grant No. 22-2-2-6-1.


\bibliographystyle{pepan}
\bibliography{TwoFieldModel}

\begin{thebibliography}{10}
\def\selectlanguageifdefined#1{
\expandafter\ifx\csname date#1\endcsname\relax
\else\selectlanguage{#1}\fi}
\providecommand*{\href}[2]{{\small #2}}
\providecommand*{\url}[1]{{\small #1}}
\providecommand*{\BibUrl}[1]{\url{#1}}
\providecommand{\BibAnnote}[1]{}
\providecommand*{\BibEmph}[1]{\emph{#1}}
\ProvideTextCommandDefault{\cyrdash}{\hbox to.8em{--\hss--}}
\providecommand*{\BibDash}{\ifdim\lastskip>0pt\unskip\nobreak\hskip.2em\fi
\cyrdash\hskip.2em\ignorespaces}

\bibitem{Salopek:1990jq}
\selectlanguageifdefined{english}
\BibEmph{Salopek D.S., Bond J.R.} {Nonlinear evolution of long wavelength
  metric fluctuations in inflationary models}~//
  \href{http://dx.doi.org/10.1103/PhysRevD.42.3936}{Phys. Rev. D}. \BibDash
\newblock 1990. \BibDash
\newblock V.~42. \BibDash
\newblock P.~3936--3962.

\bibitem{Maciejewski:2008hj}
\selectlanguageifdefined{english}
\BibEmph{Maciejewski A.J., Przybylska M., Stachowiak T., Szydlowski M.} {Global
  integrability of cosmological scalar fields}~//
  \href{http://dx.doi.org/10.1088/1751-8113/41/46/465101}{J. Phys. A}. \BibDash
\newblock 2008. \BibDash
\newblock V.~41. \BibDash
\newblock P.~465101. \BibDash
\newblock arXiv:0803.2318~[math-ph].

\bibitem{Bars:2010zh}
\selectlanguageifdefined{english}
\BibEmph{Bars I., Chen S.H.} {The Big Bang and Inflation United by an Analytic
  Solution}~// \href{http://dx.doi.org/10.1103/PhysRevD.83.043522}{Phys. Rev.
  D}. \BibDash
\newblock 2011. \BibDash
\newblock V.~83. \BibDash
\newblock P.~043522. \BibDash
\newblock arXiv:1004.0752~[hep-th].

\bibitem{Fre:2013vza}
\selectlanguageifdefined{english}
\BibEmph{Fr\'e P., Sagnotti A., Sorin A.S.} {Integrable Scalar Cosmologies I.
  Foundations and links with String Theory}~//
  \href{http://dx.doi.org/10.1016/j.nuclphysb.2013.10.015}{Nucl. Phys. B}.
  \BibDash
\newblock 2013. \BibDash
\newblock V. 877. \BibDash
\newblock P.~1028--1106. \BibDash
\newblock arXiv:1307.1910~[hep-th].

\bibitem{Kamenshchik:2013dga}
\selectlanguageifdefined{english}
\BibEmph{Kamenshchik A.Y., Pozdeeva E.O., Tronconi A., Venturi G., Vernov S.Y.}
  {Integrable cosmological models with non-minimally coupled scalar fields}~//
  \href{http://dx.doi.org/10.1088/0264-9381/31/10/105003}{Class. Quant. Grav.}
  \BibDash
\newblock 2014. \BibDash
\newblock V.~31. \BibDash
\newblock P.~105003. \BibDash
\newblock arXiv:1312.3540~[hep-th].

\bibitem{Boisseau:2015hqa}
\selectlanguageifdefined{english}
\BibEmph{Boisseau B., Giacomini H., Polarski D., Starobinsky A.A.} {Bouncing
  Universes in Scalar-Tensor Gravity Models admitting Negative Potentials}~//
  \href{http://dx.doi.org/10.1088/1475-7516/2015/07/002}{JCAP}. \BibDash
\newblock 2015. \BibDash
\newblock V.~07. \BibDash
\newblock P.~002. \BibDash
\newblock arXiv:1504.07927.

\bibitem{Kamenshchik:2015cla}
\selectlanguageifdefined{english}
\BibEmph{Kamenshchik A.Y., Pozdeeva E.O., Tronconi A., Venturi G., Vernov S.Y.}
  {Interdependence between integrable cosmological models with minimal and
  non-minimal coupling}~//
  \href{http://dx.doi.org/10.1088/0264-9381/33/1/015004}{Class. Quant. Grav.}
  \BibDash
\newblock 2016. \BibDash
\newblock V.~33, no.~1. \BibDash
\newblock P.~015004. \BibDash
\newblock arXiv:1509.00590.

\bibitem{Kamenshchik:2017ojc}
\selectlanguageifdefined{english}
\BibEmph{Kamenshchik A.Y., Pozdeeva E.O., Starobinsky A.A., Tronconi A.,
  Venturi G., Vernov S.Y.} {Induced gravity, and minimally and conformally
  coupled scalar fields in Bianchi-I cosmological models}~//
  \href{http://dx.doi.org/10.1103/PhysRevD.97.023536}{Phys. Rev. D}. \BibDash
\newblock 2018. \BibDash
\newblock V.~97, no.~2. \BibDash
\newblock P.~023536. \BibDash
\newblock arXiv:1710.02681.

\bibitem{Giacomini:2021xsx}
\selectlanguageifdefined{english}
\BibEmph{Giacomini A., Gonz\'alez E., Leon G., Paliathanasis A.} {Variational
  symmetries and superintegrability in multifield cosmology}~//
  \href{http://dx.doi.org/10.1103/PhysRevD.105.044010}{Phys. Rev. D}. \BibDash
\newblock 2022. \BibDash
\newblock V. 105, no.~4. \BibDash
\newblock P.~044010. \BibDash
\newblock arXiv:2104.13649.

\bibitem{Arefeva:2005mka}
\selectlanguageifdefined{english}
\BibEmph{Aref'eva I.Y., Koshelev A.S., Vernov S.Y.} {Crossing of the w = -1
  barrier by D3-brane dark energy model}~//
  \href{http://dx.doi.org/10.1103/PhysRevD.72.064017}{Phys. Rev. D}. \BibDash
\newblock 2005. \BibDash
\newblock V.~72. \BibDash
\newblock P.~064017. \BibDash
\newblock arXiv:astro-ph/0507067.

\bibitem{Vernov:2006dm}
\selectlanguageifdefined{english}
\BibEmph{Vernov S.Y.} {Construction of exact solutions in two-field
  cosmological models}~//
  \href{http://dx.doi.org/10.1007/s11232-008-0045-4}{Theor. Math. Phys.}
  \BibDash
\newblock 2008. \BibDash
\newblock V. 155. \BibDash
\newblock P.~544--556. \BibDash
\newblock arXiv:astro-ph/0612487.

\bibitem{Andrianov:2007ua}
\selectlanguageifdefined{english}
\BibEmph{Andrianov A.A., Cannata F., Kamenshchik A.Y., Regoli D.}
  {Reconstruction of scalar potentials in two-field cosmological models}~//
  \href{http://dx.doi.org/10.1088/1475-7516/2008/02/015}{JCAP}. \BibDash
\newblock 2008. \BibDash
\newblock V.~02. \BibDash
\newblock P.~015. \BibDash
\newblock arXiv:0711.4300~[gr-qc].

\bibitem{Arefeva:2009tkq}
\selectlanguageifdefined{english}
\BibEmph{Aref'eva I.Y., Bulatov N.V., Vernov S.Y.} {Stable Exact Solutions in
  Cosmological Models with Two Scalar Fields}~//
  \href{http://dx.doi.org/10.1007/s11232-010-0063-x}{Theor. Math. Phys.}
  \BibDash
\newblock 2010. \BibDash
\newblock V. 163. \BibDash
\newblock P.~788--803. \BibDash
\newblock arXiv:0911.5105~[hep-th].

\bibitem{Paliathanasis:2018vru}
\selectlanguageifdefined{english}
\BibEmph{Paliathanasis A., Leon G., Pan S.} {Exact Solutions in Chiral
  Cosmology}~// \href{http://dx.doi.org/10.1007/s10714-019-2594-2}{Gen. Rel.
  Grav.} \BibDash
\newblock 2019. \BibDash
\newblock V.~51, no.~9. \BibDash
\newblock P.~106. \BibDash
\newblock arXiv:1811.10038.

\bibitem{Chervon:2019nwq}
\selectlanguageifdefined{english}
\BibEmph{Chervon S.V., Fomin I.V., Pozdeeva E.O., Sami M., Vernov S.Y.}
  {Superpotential method for chiral cosmological models connected with modified
  gravity}~// \href{http://dx.doi.org/10.1103/PhysRevD.100.063522}{Phys. Rev.
  D}. \BibDash
\newblock 2019. \BibDash
\newblock V. 100, no.~6. \BibDash
\newblock P.~063522. \BibDash
\newblock arXiv:1904.11264.

\bibitem{Faraoni:2021opj}
\selectlanguageifdefined{english}
\BibEmph{Faraoni V., Jose S., Dussault S.} {Multi-fluid cosmology in Einstein
  gravity: analytical solutions}~//
  \href{http://dx.doi.org/10.1007/s10714-021-02879-z}{Gen. Rel. Grav.} \BibDash
\newblock 2021. \BibDash
\newblock V.~53, no.~12. \BibDash
\newblock P.~109. \BibDash
\newblock arXiv:2107.12488.

\bibitem{Fomin:2021snm}
\selectlanguageifdefined{english}
\BibEmph{Fomin I.V., Chervon S.V.} {New method of exponential potentials
  reconstruction based on given scale factor in phantonical two-field
  models}~// \href{http://dx.doi.org/10.1088/1475-7516/2022/04/025}{JCAP}.
  \BibDash
\newblock 2022. \BibDash
\newblock V.~04, no.~04. \BibDash
\newblock P.~025. \BibDash
\newblock arXiv:2112.09359.

\bibitem{Mitsopoulos:2021qpu}
\selectlanguageifdefined{english}
\BibEmph{Mitsopoulos A., Tsamparlis M., Leon G., Paliathanasis A.} {New
  Conservation Laws and Exact Cosmological Solutions in Brans\textendash{}Dicke
  Cosmology with an Extra Scalar Field}~//
  \href{http://dx.doi.org/10.3390/sym13081364}{Symmetry}. \BibDash
\newblock 2021. \BibDash
\newblock V.~13, no.~8. \BibDash
\newblock P.~1364. \BibDash
\newblock arXiv:2105.02766.

\bibitem{Russo:2023nir}
\selectlanguageifdefined{english}
\BibEmph{Russo J.G.} {New exact solutions in multi-scalar field cosmology}~//
  \href{http://dx.doi.org/10.1088/1475-7516/2023/07/066}{JCAP}. \BibDash
\newblock 2023. \BibDash
\newblock V.~07. \BibDash
\newblock P.~066. \BibDash
\newblock arXiv:2304.12360.

\bibitem{Kamenshchik:2016atu}
\selectlanguageifdefined{english}
\BibEmph{Kamenshchik A.Y., Pozdeeva E.O., Tronconi A., Venturi G., Vernov S.Y.}
  {Integrable cosmological models in the Einstein and in the Jordan frames and
  Bianchi-I cosmology}~//
  \href{http://dx.doi.org/10.1134/S1063779618010173}{Phys. Part. Nucl.}
  \BibDash
\newblock 2018. \BibDash
\newblock V.~49, no.~1. \BibDash
\newblock P.~1--4. \BibDash
\newblock arXiv:1606.04260.

\bibitem{Kaiser:2010ps}
\selectlanguageifdefined{english}
\BibEmph{Kaiser D.I.} {Conformal Transformations with Multiple Scalar
  Fields}~// \href{http://dx.doi.org/10.1103/PhysRevD.81.084044}{Phys. Rev. D}.
  \BibDash
\newblock 2010. \BibDash
\newblock V.~81. \BibDash
\newblock P.~084044. \BibDash
\newblock arXiv:1003.1159~[gr-qc].

\bibitem{Chervon:1995jx}
\selectlanguageifdefined{english}
\BibEmph{Chervon S.V.} {On the chiral model of cosmological inflation}~//
  \href{http://dx.doi.org/10.1007/BF00559313}{Russ. Phys. J.} \BibDash
\newblock 1995. \BibDash
\newblock V.~38. \BibDash
\newblock P.~539--543.

\bibitem{Chervon:2013btx}
\selectlanguageifdefined{english}
\BibEmph{Chervon S.V.} {Chiral Cosmological Models: Dark Sector Fields
  Description}~// Quant. Matt. \BibDash
\newblock 2013. \BibDash
\newblock V.~2. \BibDash
\newblock P.~71--82. \BibDash
\newblock arXiv:1403.7452~[gr-qc].

\bibitem{Ivanov:2021ily}
\selectlanguageifdefined{english}
\BibEmph{Ivanov V.R., Vernov S.Y.} {Integrable cosmological models with an
  additional scalar field}~//
  \href{http://dx.doi.org/10.1140/epjc/s10052-021-09792-4}{Eur. Phys. J. C}.
  \BibDash
\newblock 2021. \BibDash
\newblock V.~81, no.~11. \BibDash
\newblock P.~985. \BibDash
\newblock arXiv:2108.10276.

\bibitem{Ivanov:2024nnd}
\selectlanguageifdefined{english}
\BibEmph{Ivanov V.R., Vernov S.Y.} {New Integrable Chiral Cosmological Models
  with Two Scalar Fields}~// Phys. Rev. D. \BibDash
\newblock 2024. \BibDash
\newblock V. 110 to be published. \BibDash
\newblock arXiv:2407.12732.

\bibitem{Bars:2011mh}
\selectlanguageifdefined{english}
\BibEmph{Bars I., Chen S.H., Turok N.} {Geodesically Complete Analytic
  Solutions for a Cyclic Universe}~//
  \href{http://dx.doi.org/10.1103/PhysRevD.84.083513}{Phys. Rev. D}. \BibDash
\newblock 2011. \BibDash
\newblock V.~84. \BibDash
\newblock P.~083513. \BibDash
\newblock arXiv:1105.3606~[hep-th].

\bibitem{Ivanov:2023uvh}
\selectlanguageifdefined{english}
\BibEmph{Ivanov V.R., Vernov S.Y.} {Anisotropic Solutions for
  $\boldsymbol{R^{2}}$ Gravity Model with a Scalar Field}~//
  \href{http://dx.doi.org/10.1134/S1063778824010204}{Phys. Atom. Nucl.}
  \BibDash
\newblock 2023. \BibDash
\newblock V.~86, no.~6. \BibDash
\newblock P.~1526--1532. \BibDash
\newblock arXiv:2301.06836.

\end{thebibliography}

\end{document}